# Transport and Spectroscopic Studies of the Effects of Fullerene Structure on the Efficiency and Lifetime of Polythiophene-based Solar Cells

Emilee L. Sena[a,b], Justin H. Peel[a], Devin Wesenberg[a], Shreya Nathan[a], Marianne Wallis[b], Maxwell J. Giammona[b], Thorsteinn Adalsteinsson[b], Brian J. McNelis[b], and Richard P. Barber, Jr.[a,*]

[a]Department of Physics, Santa Clara University, Santa Clara, CA 95053
[b]Department of Chemistry and Biochemistry, Santa Clara University, Santa Clara, CA 95053

**Abstract**

Time-dependent measurements of both power conversion efficiency and ultraviolet-visible absorption spectroscopy have been observed for solar cell blends containing the polymer poly(3-hexylthiophene-2,5-diyl) (P3HT) with two different functionalized $C_{60}$ electron acceptor molecules: commercially available [6,6]-phenyl $C_{61}$ butyric acid methyl ester (PCBM) or [6,6]-phenyl $C_{61}$ butyric acid octadecyl ester (PCBOD) produced in this laboratory. Efficiency was found to decay with an exponential time dependence, while spectroscopic features show saturating exponential behavior. Time constants extracted from both types of measurements showed reasonable agreement for samples produced from the same blend. In comparison to the PCBM samples, the stability of the PCBOD blends was significantly enhanced, while both absorption and power conversion efficiency were decreased.

*Corresponding author. Tel.: +1 408-554-4315; fax +1 408-554-6965
E-mail address: rbarber@scu.edu (R. P. Barber, Jr.)

# 1. Introduction

Polymer photovoltaics offer the promise of an inexpensive and mechanically robust alternative to established solar cell technologies. Recent progress has seen the establishment of manufacturing approaches [1] and improvement in both power conversion efficiency [2, 3] and stability [4, 5]. Despite these gains there remain many unanswered questions regarding the composition and precise mechanisms that govern the performance of these devices [6]. To date the primary research focus has been on improving power conversion efficiency, with significantly less attention on characterizing stability and understanding the mechanisms that limit the lifetime of these materials [7]. Furthermore, the published work in this field typically does not present results from experiments which sample from the large parameter space of materials and device preparation. Publication of the results and discussion of only the systems that simply follow empirical recipes for fullerene and polymer ratios and fabrication parameters is insufficient to determine the underlying mechanisms that influence device performance. In order to develop a more robust methodology and process for evaluating solar cell composition and fabrication parameters as they affect device performance, we present coordinated results from transport and optical measurements of samples based on the electron donor Poly(3-hexylthiophene-2,5-diyl) (P3HT). To elucidate the mechanisms that control efficiency *and* lifetime, we repeat these measurements on each sample over time scales of a few hours to weeks while changing only one parameter at a time. This effort is intended to produce a broader and more systematic data set.

Previous work has shown that the addition of functionalized fullerenes enhanced device lifetime for devices based on Poly[2-methoxy-5-(2-ethylhexyloxy)-1,4-phenylenevinylene] (MEH-PPV) as the electron donor [8]. Our recent results both corroborate that conclusion and indicate an additional improvement by varying the structure of the functionalized fullerene [9]. Specifically, we have found significant lifetime improvements by using [6,6]-Phenyl $C_{61}$ butyric acid octadecyl ester (PCBOD) in place of the commercially available electron acceptor PCBM. Coupled with this increase in device lifetime, however, is a reduction in power conversion efficiency. In this current study we have extended our results to the more commonly studied P3HT system. Fig. 1 compares the structure of PCBOD and the commonly used [6,6]-Phenyl $C_{61}$ butyric acid methyl ester (PCBM).

In order to compare both the PCBOD and PCBM systems, we have adopted the molar fraction $x$ to describe the relative content of our polymer fullerene blends.

$$x = \frac{m_{C60} / MW_{C60}}{m_{C60} / MW_{C60} + m_{P3HT} / MW_{P3HT}}$$

with $MW$ denoting the molecular weight of each species and $m$ denoting its mass. Given that the degree of polymerization can vary, we use the molar mass of the P3HT monomer. In this description, a mole fraction of $x=0.5$ would represent a blend which has one $C_{60}$ molecule per P3HT monomer. The standard widely used in the literature is equal weight concentrations of

PCBM:P3HT, which corresponds to x = 0.16.  While the equal-weight ratio might serve as a simple starting point for initial experiments using new materials, it is at best cumbersome and perhaps even nonsensical when we compare fullerene-polymer blends using *different* fullerenes with *different* molecular weights.  It is well-established that the functionalized fullerene is the active component in these additives, and the functionalization increases solubility to improve device processing [10].  In previous work [2, 9, 11-15] additional structural changes have been made which dramatically change molecular weight.  Our strong opinion is that we should be using mole fraction to clearly establish the ratio of fundamental units: fullerene-to-monomer units in the polymer.  For our work, we want a direct comparison between the PCBM- and PCBOD-based devices that have some systematic link and make sense chemically; therefore we use molar fraction.

Important gains in efficiency have been observed using $C_{70}$-substituted PCBM [2]. However this molecule is more than 10% heavier than PCBM.  Although weight ratios can be converted, direct comparison would be more straightforward if mole fraction were adopted.   An example of the beneficial use of mole fraction appears in models for the crystal structure of these blends that clearly depict one fullerene to one monomer [16].  This ratio represents a mole fraction of *x=0.5* within the domains and not the macroscopic value *x=0.16* equivalent to the equal-weight recipe.  The use of mole fraction in that work significantly enhances the clarity of the results.  The difference in these values can be understood within the context of recent energy-filtered transmission electron microscopy images which have revealed the P3HT-rich and PCBM-rich domains within the active layers [17].  This result can accommodate both the crystalline model

($x=0.5$ locally) [16, 18] and other studies of the morphology and diffusion within these samples ($x=0.16$ globally) [6, 19, 20].

## 2. Experimental

Solution processed samples were prepared inside an inert-atmosphere glove box. Indium-tin-oxide (ITO) pre-coated glass substrates were patterned, cleaned and then spin-coated with poly[3,4-ethylenedioxythiophene]:poly[styrenesulfonate] (PEDOT:PSS) and baked at 200 °C for two hours. The active layer blend of P3HT with either PCBM or PCBOD was then spin cast at roughly 1000 rpm from chlorobenzene solutions. Active-layer solution weight fractions were roughly 1.5%. We expect these parameters to produce active layer thicknesses of about 150 nm based on previous experience [21], but we have not performed direct measurements on these samples. Interestingly we do find that PCBOD-based samples at the same weight percent produce optically thinner layers than those with PCBM. The result is perhaps expected given the higher solubility of PCBOD in chlorobenzene. PCBM is the electron acceptor widely used in polymer photovoltaic measurements and was acquired commercially [22]. PCBOD was synthesized in our laboratories from coupling [6,6]-phenyl $C_{61}$ butyric acid with octadecanol as discussed previously [9]. After spin casting, samples were annealed on a hot plate inside the glove box for one hour at temperatures between 40 and 340 °C. In some cases, samples were not annealed prior to further processing. During our initial experiments it was determined that the actual sample temperature during annealing was lower than that of the hot plate surface. In order to correct for this difference, we have calibrated the hot plate/sample system using a 0.25 mm type-E thermocouple placed on the sample surface (top surface of the glass) during heating. We found that sample temperatures were roughly 20 ºC lower when the hotplate temperature was about 150 ºC and nearly 40 ºC lower at 250 ºC.  Samples that were used for UV-vis spectroscopy studies were prepared by the same method, but microscope slides replaced the patterned ITO substrates and the samples were ready for measurement after spin-casting the active layer and annealing.  Samples to be used for electrical transport characterization were transferred to a standard bell-jar evaporator system equipped with a quartz crystal thickness monitor where they were finished by evaporating ~1 nm of LiF followed by 100 nm of Al to form the top electrodes.  As the evaporator is not integrated into the glove box, samples were transferred between the two using a vacuum tight vessel carrying dry nitrogen atmosphere. The elapsed time that samples were exposed to ambient air was typically under 5 minutes.  A schematic of the transport sample structure is shown in Fig. 1.

Both electrical transport and spectroscopic measurements were conducted in ambient atmosphere immediately after completing the sample preparation.  Current-voltage (*I-V*) transport characteristics of the devices were measured alternately in darkness and illuminated by a PV Measurements, Inc. Small-Area Class-B Solar Simulator.  Automated transport data collection utilized a MATLAB controlled routine via an IEEE 488 Bus interfaced Keithley 2400 SourceMeter.  UV-vis measurements were performed using a Varian Cary 50 spectrometer.  Samples were measured over periods of hours or days for both spectroscopic and transport experiments.  Illumination was used only during the actual measurements; otherwise the device remained in ambient but low light conditions.  The laboratory temperature was controlled at 23-26 ºC with a relative humidity range of 50-70% (not directly controlled).

## 3. Results and Discussion

Fig 2 shows a typical set of transport curves. *I-V* measurements are normally recorded in 10-30 minute intervals depending on the apparent rate of change, and many interleaving curves were removed from this figure for clarity. This particular sample was a PCBM blend, however PCBOD samples yielded similar results, albeit at lower current scales. The inset of this figure shows the calculated power conversion efficiency as a function of time η(t). As η(t) appears linear in a semilog plot, it is apparent that it follows an exponential decay

$$\eta(t) = \eta(0)\exp(-t/\tau),$$

where a characteristic (1/e) lifetime τ can be derived from the slope (-1/τ) [9]. Such a slope is shown in the Fig. 2 inset. In cases that η(t) does not strictly follow an exponential decay, we have chosen to simply find the time at which η drops to 1/e of its initial value. This approach allows us to derive a figure of merit with which to compare the lifetime of various sample preparations. Specifically, we occasionally observe samples that first improve in efficiency before the onset of degradation. In some cases these samples then decay exponentially and we report both τ and 1/e time.

Fig. 3 presents results for PCBM:P3HT samples for two different blends x=0.16 (standard equal weight blend) and x=0.40 (the optimum result for the PCBOD samples to be discussed). The top frame shows the initial power conversion efficiency as a function of anneal temperature. The lower frame displays the degradation times for the corresponding samples. τ values as discussed above are shown as open symbols and filled symbols denote 1/e times.  These results do not appear to favor the x=0.16 samples with respect to power conversion efficiency, however the device lifetimes are clearly better for these samples when produced at the higher anneal temperatures.  It is also important to note that we have focused on systematic sample preparation for these studies.  We have not focused on optimizing PCBM:P3HT devices and as such we do not obtain (or expect) state-of-the-art efficiencies.

Since PCBOD:P3HT blend devices have not been previously reported in the literature and the structure and molecular weight of PCBOD is significantly different than PCBM, it was necessary to determine reasonable blend and annealing parameters for making devices.  Fig. 4 shows results on unannealed samples using a range of stoichiometry.  The same molar fraction blend as PCBM (x=0.16) showed rather poor results.  By far the best efficiencies (roughly 0.1%) for as-spun devices were produced by molar fractions near 0.4.  Without more comprehensive testing, it is impossible to know whether this blend is also optimum for annealed samples, but that endeavor presents a dramatically larger parameter space.

In Fig 5 we show the effect of annealing on PCBOD:P3HT samples with x=0.4 as a function of anneal temperature.  Again we show initial power conversion efficiency $\eta(0)$ in the upper frame and the lifetime in the lower one.  In comparing these results with those for PCBM:P3HT blends at x=0.16 in Fig. 3, we note that these latter samples have much higher efficiencies, but the PCBOD system still shows better lifetimes.  In general, we see that the rate of degradation is lower for PCBOD samples while the PCBM devices show a higher efficiency.  It is also worth noting that for both fullerenes the degradation is typically exponential. This result is in contrast to earlier work using phenylenevinylene-based devices [9].

An important component of our research is the coordinated measurement of both transport and spectroscopic data for samples produced from the same blend solutions. Furthermore these experiments extend to time dependences for both measurements as we focus on characterizing device degradation and investigating its mechanisms. Despite the large differences in power conversion efficiency and lifetime, previous studies show that UV-vis absorbance spectra vary little between annealed and unannealed samples [23]. However, spectral changes were not studied over an extended length of time where differential measurements show significant changes in our spectroscopic studies. We show that as samples degrade, the characteristic absorbance spectra for unannealed and annealed samples evolve. In general we observe different characteristic times for different spectral peaks with unique signatures that distinguish the behavior of the PCBM:P3HT and PCBOD:P3HT samples. PCBM:P3HT samples show decreasing absorbance over time in the range of regioregular P3HT's maximum absorbance, approximately 450 to 600 nm [24]. The opposite is true for the PCBOD:P3HT blend. It is perhaps noteworthy that these trends were reversed in the MEH-PPV system studied previously [9].

The lower frames of Figs. 6 and 7 show examples of differential spectra for PCBM and PCBOD samples respectively with vertical arrows showing the direction of change as time increases. Each line on the figures represents the difference between a measurement at specified time (t) and the initial measurement. Positive absorbance values indicate an increase in absorbance and negative values indicate a decrease. The upper frames depict the initial and final raw spectra for these samples.

Thin films composed of regioregular P3HT and PCBM absorb the most UV and visible light between 450 and 600 nm, a range which includes the maximum absorbance peaks of pure P3HT thin films [25]. Our UV-vis results for PCBM:P3HT samples are consistent with these literature values. PCBOD:P3HT thin films do not absorb most strongly in this range, but exhibit significant spectral changes within it. The time-dependent changes within this range suggest that the polymer is changing absorptivity (and likewise device efficiency) over time, likely due to some molecular reorganization. The following results focus solely on changes within the 450 to 600 nm range.

PCBM:P3HT samples (x=0.16, both unannealed and annealed) demonstrate time-dependent changes in absorbance levels, but no red- or blue-shifts (Fig. 6). The absorbance at 550 nm decreased, with the largest changes in samples annealed at temperatures up to 195 ºC. Above 195 ºC, the trend was less significant. This same trend was apparent, yet larger, in samples with greater PCBM content (x=0.4).

PCBOD:P3HT samples (x=0.4, annealed at 195 ºC) showed red-shifts and an increase in absorbance (Fig. 7). The spectra initially displayed a shoulder at 515 nm, but then a shoulder at 550 nm developed and grew much larger than the 515 nm shoulder. This growth saturated after about one week. All other PCBOD:P3HT samples (x=0.4) gave the same general trend, and the increase in absorbance at 550 nm was more significant in unannealed samples. Negligible changes occurred when samples were annealed past 218 ºC, P3HT's melting point [26].

In order to derive values for characteristic times from the spectroscopic data, we use our previous approach [9]. Typically we see peaks in the time-dependent spectra which grow and saturate. Such temporal behavior is often consistent with a saturating exponential. As such we can model this behavior as

$$abs(t) = abs(\infty)[1 - \exp(-t/\tau)]$$

where $abs(\infty)$ denotes the differential absorbance after long times (the saturated value). As before, in cases when the exponential saturation is not as easily fitted, we look for the [1-1/e] saturation time to yield a characteristic time scale. Such a fit is shown in the inset of Fig. 7 for the 551 nm peak. That peak grows with time and its value fits well to a saturating exponential with final value 0.018 and time constant of 1.69 days (roughly 2400 minutes). Time constants gathered from various annealing temperatures and stoichiometries are shown in Fig. 8. The top figure demonstrates that τ for PCBM:P3HT decreased with annealing temperature. In contrast, in the bottom figure, PCBOD:P3HT samples yielded a maximum time constant at 195 ºC.

From these data, we have developed a model to explain device performance and spectral trends and the correlation of their time dependences. Increasing the ratio of fullerene (C60 or PCBM) additives has been shown to decrease the film absorption between 450 and 600 nm [25]. This range relates to absorption maximum for P3HT and corresponds to a highly ordered, crystalline P3HT film which is required for efficient devices [24, 27]. Therefore, the weak absorbance in this region of our PCBOD:P3HT films is a clear indication that these samples will have lower efficiency, as reflected in our device studies. This result suggests that the C18 chain on the PCBOD interferes with the crystallization of the P3HT. Although the relatively slow spectral red-shifts observed in PCBOD:P3HT might indicate some increase in polymer order [25], this change is likely small compared to the disruption in the ordering caused by the chains. We observe clear processing differences between the PCBM and PCBOD blends at the same weight concentration of solution. We find that PCBOD's dramatically increased solubility makes the spin-cast films much thinner. In fact, samples which can barely be seen as coating the substrate with the naked eye produce 0.1% efficient devices. The C18 seems to make the fullerene more miscible with P3HT lowering its ability to organize in crystalline domains and it surfactant-like structure decreases the viscosity of the solution resulting in thinner devices.

The annealing and performance experiments also follow an understandable and consistent trend. As spun devices adopt a kinetically-frozen organization in the film and the two components separate into phases and self-organize into local minima when the film is produced. [28] Annealing allows for reordering of the components of the film and the predominant change is the increase in the organization of the P3HT. PCBM has been shown to have significant mobility in P3HT films on annealing, [6] and it reasonable to predict that PCBOD would not have similar mobility. In fact, this was our working model to increase stability of devices: that the rate of PCBM-based device degradation was consistent with reorganization and both our current study and previous work [9] support this assertion. However, our current results show that at about the same annealing temperature, PCBM devices are 10-fold more efficient, suggesting that fullerene mobility facilitates the P3HT reorganization to prepare a higher efficiency device. Therefore, in the extreme cases as it relates to fullerene ester substitution (methyl vs. octadecyl), stability and efficiency constitute a trade-off in these device performance characteristics. Fortunately for our future work, there is great variability in the structure of the alkyl substituent still to be explored that could enhance both performance *and* lifetime. Preliminary results from samples made with [6,6]-Phenyl $C_{61}$ butyric acid octyl ester (PCBO) exhibit efficiencies that are improved by a factor of four over the PCBOD results. We are exploring other fullerene esters that will be compatible with or enhance the crystallization of P3HT and reinforce the required fullerene domains to prepare efficient, robust devices.

## 4. Conclusions

Using time dependent measurements of device efficiency and differential UV-vis spectroscopy, we have demonstrated that varying the chain on the fullerene ester can affect the lifetime and efficiency of P3HT:fullerene based solar cells. Substituting octadecyl for methyl in the fullerene yielded 3-fold improved lifetimes, however the PCBOD devices were much thinner than PCBM devices and 10-fold less efficient. Although the device thickness could be contributing to the lower efficiencies, is it clear from spectroscopic and device measurements that the C18 chain affects the crystallinity of the P3HT thereby lowering the transport properties of the devices. Preliminary studies have shown that a shorter octyl chain is only 5-fold less efficient, however this system has not yet been optimized. Future work will continue to focus on wide ranging possibilities in ester chain variations that could lead to improved device lifetime and performance characteristics. These simple changes in fullerene structure provide fundamental understanding of the device dynamics and if successful, the synthetic targets are amenable to commercial applications.

## 5. Acknowledgements

This project has been supported by Santa Clara University funding including a Santa Clara University's BIN-REU: UCSC BIN-RDI, NASA Grant NNX09AQ44A; a Center for Science Technology and Society Grant; and a Center for Science Technology and Society Roelandts Fellowship.  Additional funding was provided by a grant from IntelliVision Technologies.  We acknowledge valuable discussions with D. Romero.

## 6. Figure Captions

Fig. 1. Electron acceptor molecules and sample layout used in this study: a) [6,6]-phenyl $C_{61}$ butyric acid methyl ester (PCBM), b) [6,6]-phenyl $C_{61}$ butyric acid octadecyl ester (PCBOD), and c) schematic of solar cell layers.

Fig. 2. Typical current-voltage (*I-V*) curves showing the degradation of a PCBM:P3HT device in ambient conditions. The arrow indicates the progression of time. *Inset:* a semilog plot of the power conversion efficiency of this device as a function of time $\eta$ in ambient conditions. The solid line fit shows the slope used to extract the characteristic time $\tau$.

Fig. 3. Data for two series of PCBM:P3HT devices (molar fraction 0.16 and 0.40). We plot a) the initial power conversion efficiency $\eta$ and b) the decay time constant $\tau$ as a function of anneal temperature. The gray-filled symbols indicate samples for which we calculated 1/e times (see manuscript).

Fig. 4. Initial power conversion efficiency $\eta$ for PCBOD:P3HT devices as a function of PCBOD molar fraction. Note the maximum near 0.4 molar fraction.

Fig. 5. Data for five series of PCBOD:P3HT devices (all approximately 0.4 molar fraction). We plot a) the initial power conversion efficiency $\eta$ and b) the decay time constant $\tau$ as a function of anneal temperature. The gray-filled symbols indicate samples for which we calculated 1/e times (see manuscript).

Fig. 6. a) Initial and final UV-vis absorption spectrum for an x = 0.16 PCBM:P3HT device. b) Differential UV-vis absorption spectrum ($\Delta abs.$) for this same device evolving with time (denoted by the arrow).

Fig. 7. a) Initial and final UV-vis absorption spectrum for an x = 0.4 PCBOD:P3HT device. b) Differential UV-vis absorption spectrum ($\Delta abs.$) for this same device evolving with time (denoted by the arrow).

Fig. 8. Saturation time constants $\tau$ as derived from various spectral peaks in the UV-vis absorption data. a) PCBM:P3HT results for four different molar fractions. b) PCBOD:P3HT results for x = 0.4. Note that the longest time constants for PCBOD-based devices occur for the 195 °C annealed samples, consistent with the transport results shown in Fig. 5.

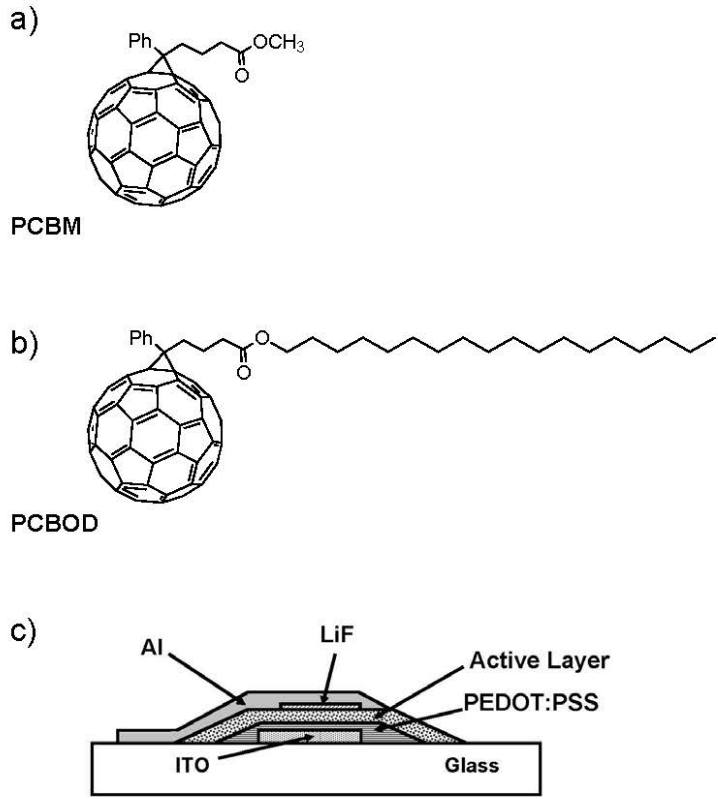

Fig 1

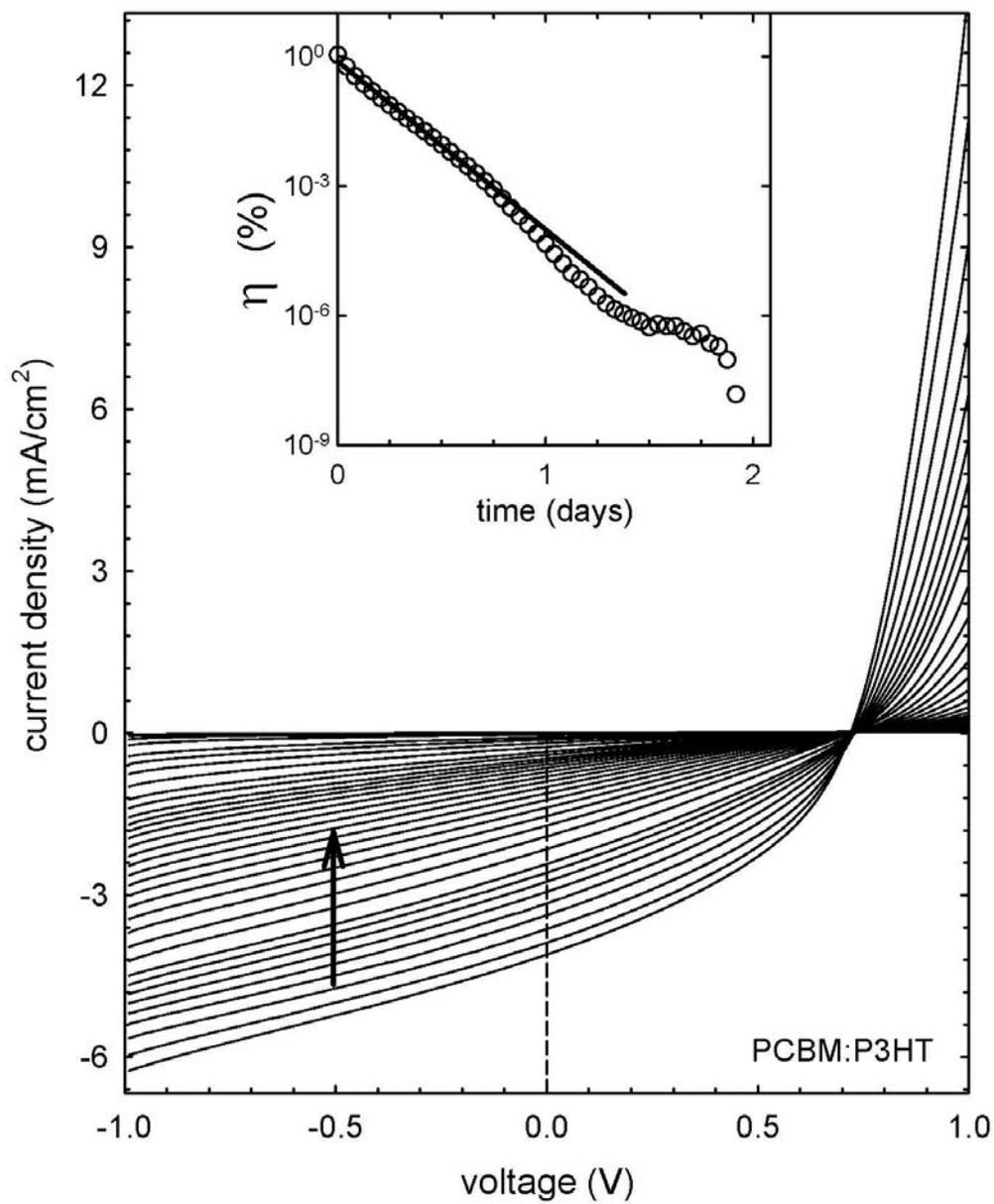

Fig 2

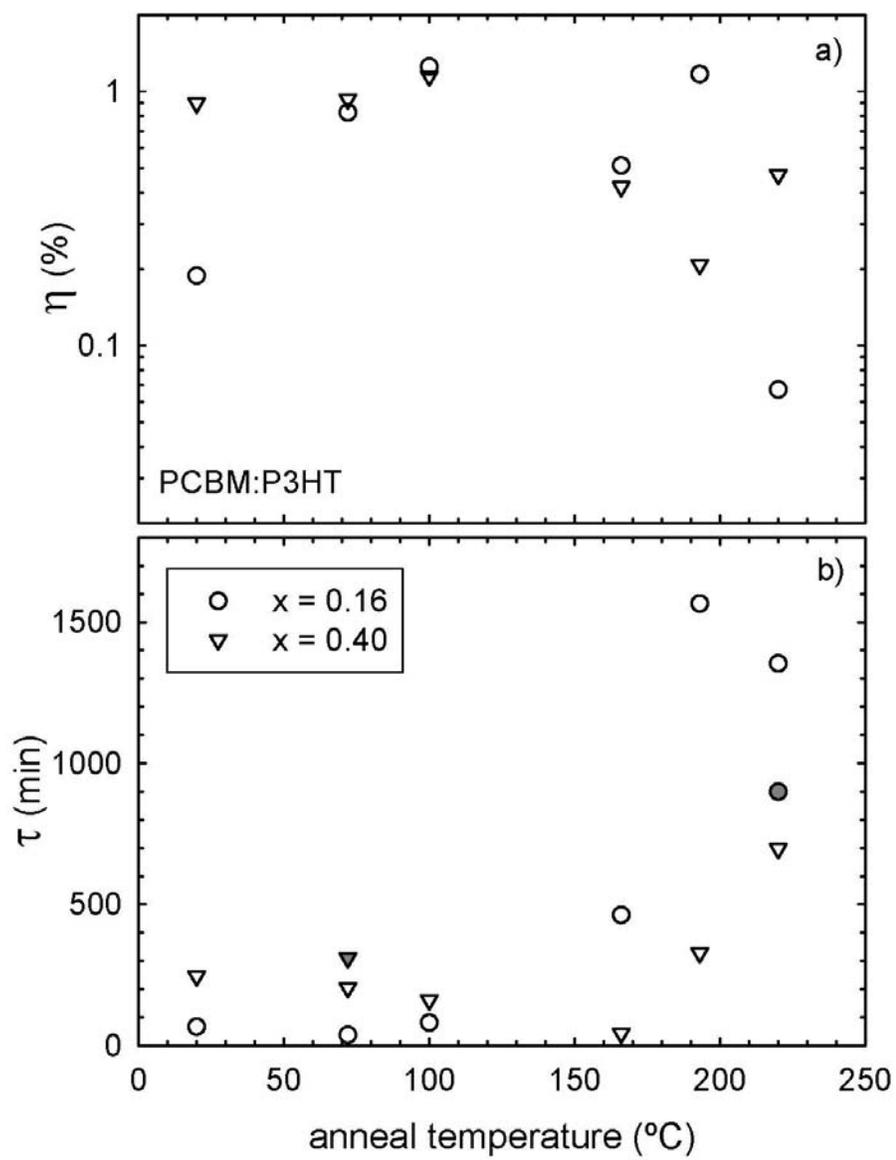

Fig 3

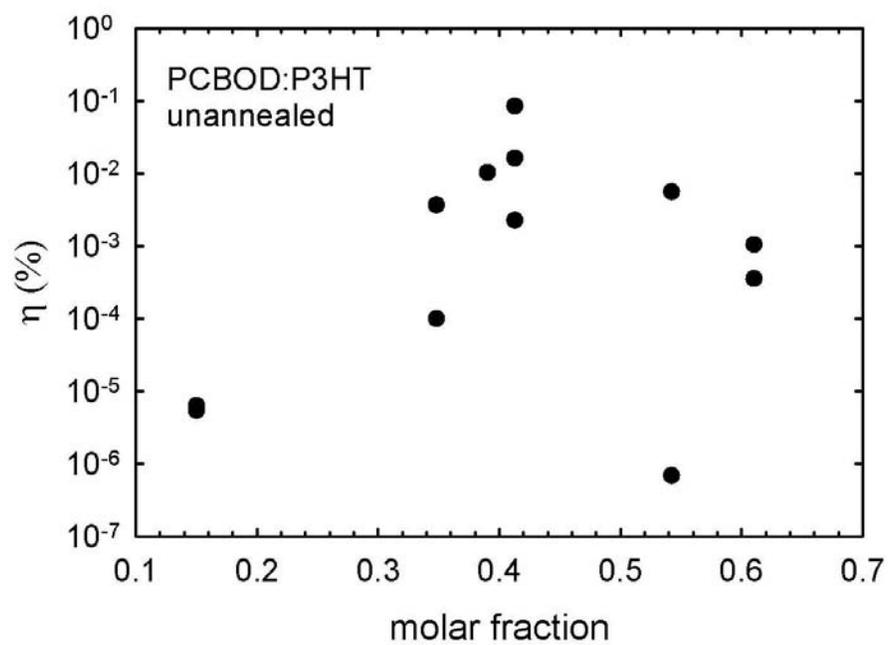

Fig 4

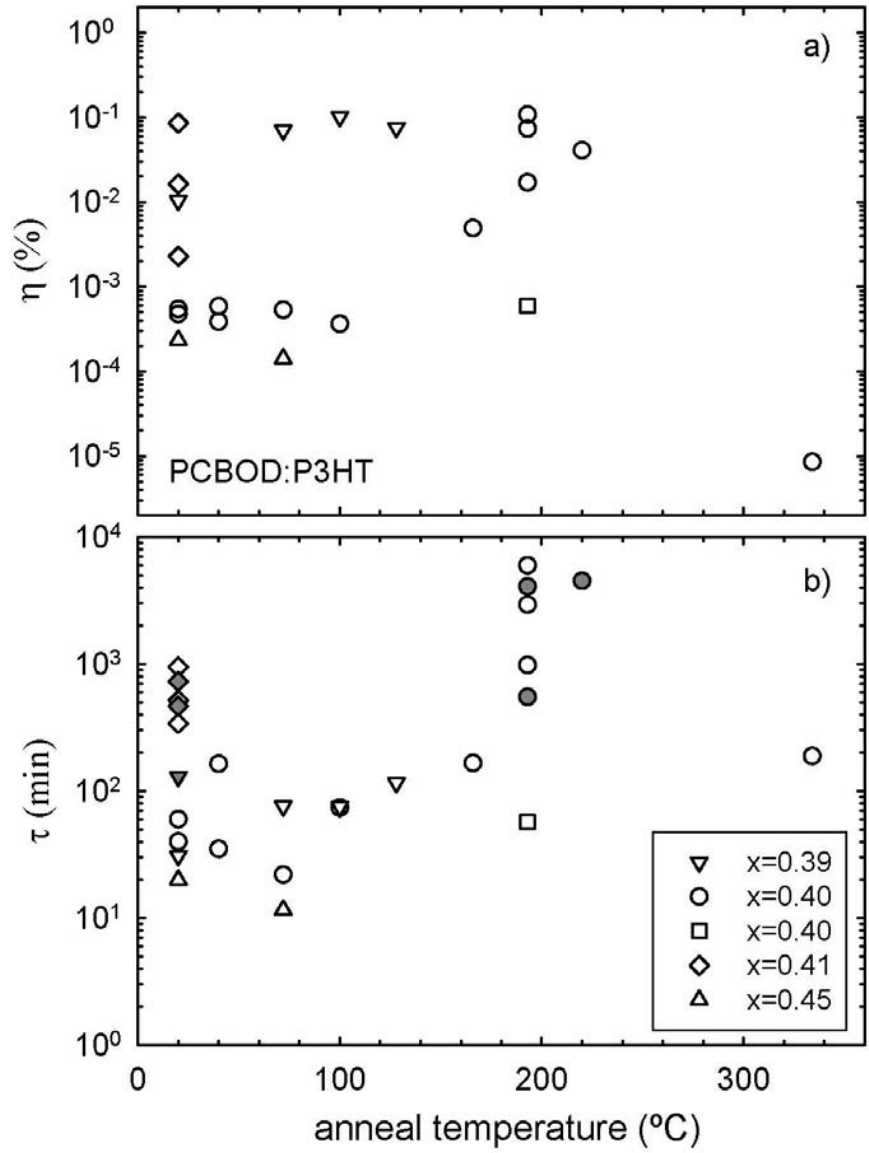

Fig 5

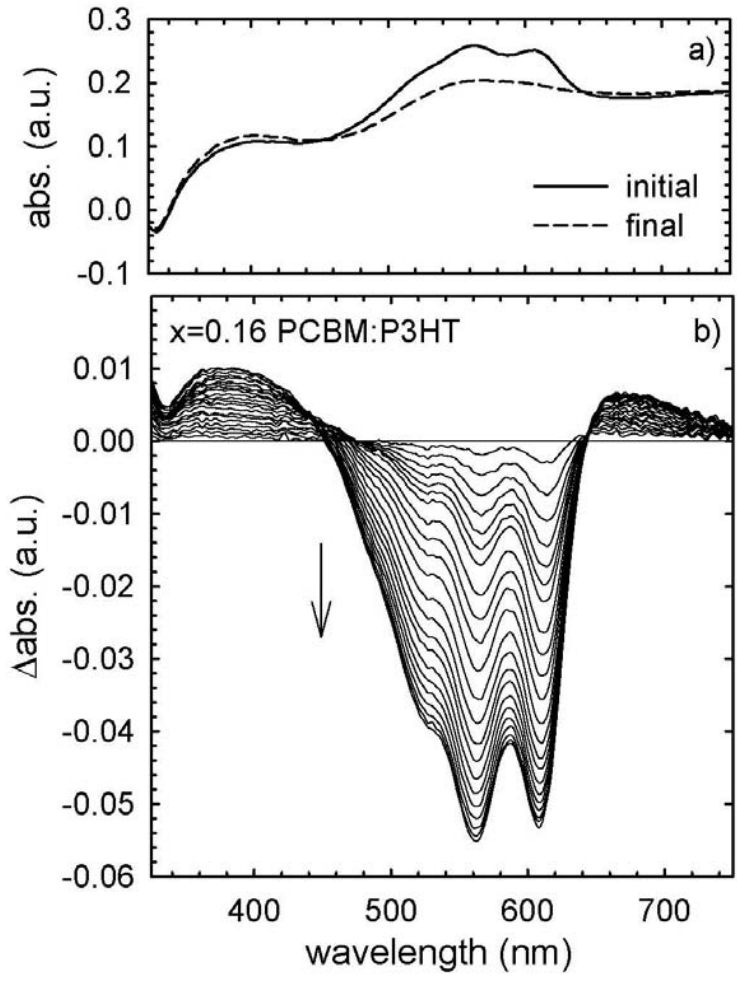

Fig 6

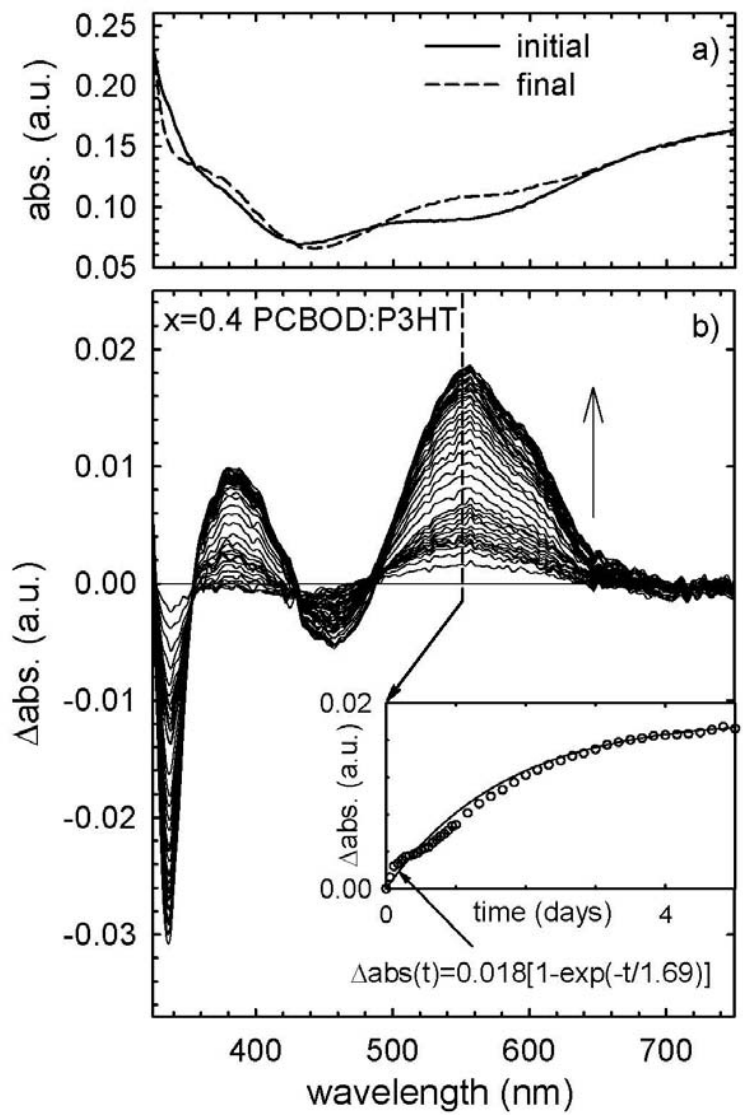

Fig 7

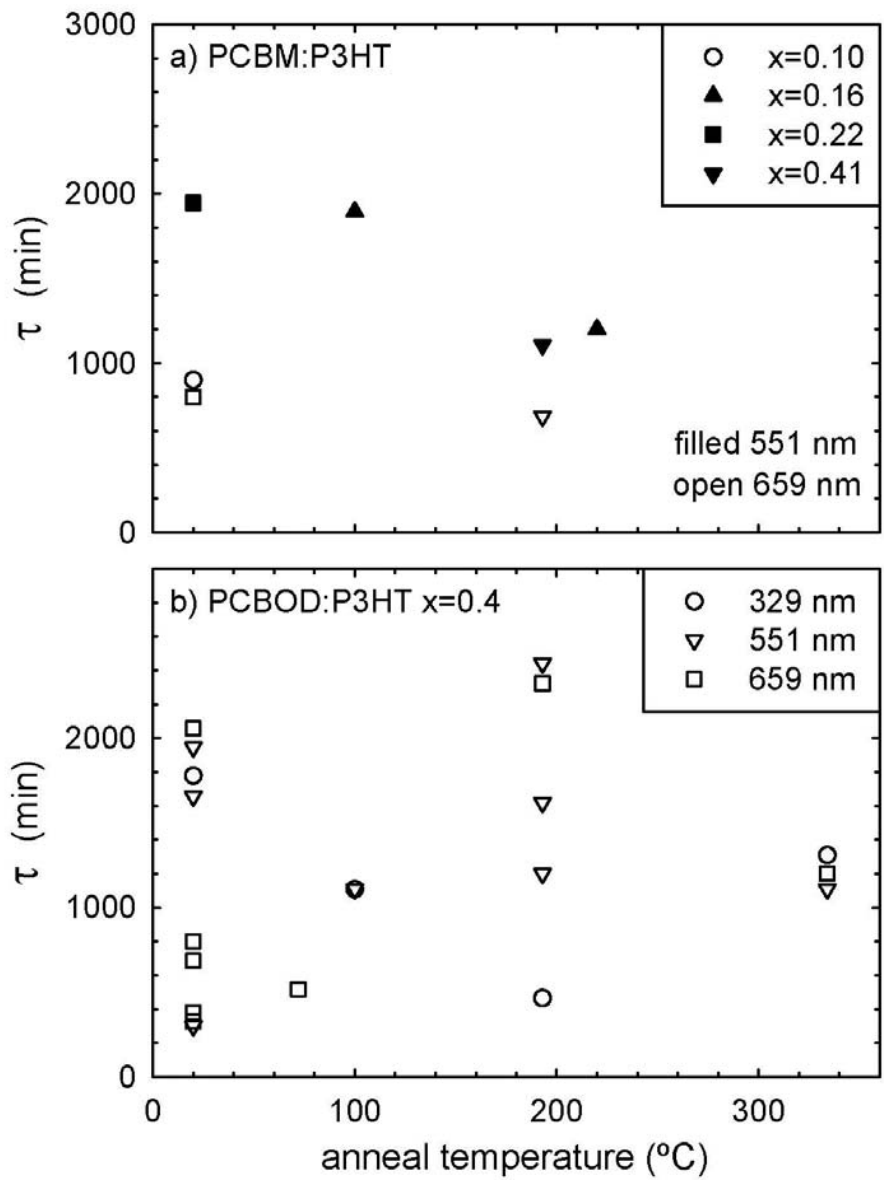

Fig 8